# Belief in False Information: A Human-Centered Security Risk in Sociotechnical Systems

**Fabian Walke**
Heilbronn University

**Thaddäa Nürnberger**
University of Hagen

### Abstract

*This paper provides a comprehensive literature review on the belief in false information, including misinformation, disinformation, and fake information. It addresses the increasing societal concern regarding false information, which is fueled by technological progress, especially advancements in artificial intelligence. This review systematically identifies and categorizes factors that influence the belief in false information. The review identifies 24 influence factors grouped into six main categories: demographic factors, personality traits, psychological factors, policy and values, media consumption, and preventive factors. Key findings highlight that lower education levels, high extraversion, low agreeableness, high neuroticism, and low cognitive reflection significantly increase belief in false information. The effectiveness of preventive strategies like labeling false information and promoting reflection about correctness is also discussed. This literature review conceptualizes belief in false information as a human-centered security risk in sociotechnical systems, as it can be exploited to manipulate decisions, undermine trust, and increase susceptibility to social engineering. It aims to inform preventive strategies that strengthen socio-technical security and societal resilience.*

## 1. Introduction

The impact of false information (FI) extends beyond individuals, affecting society as a whole. Although false information is not new, it is currently perceived as a major problem (Osmundsen et al., 2021). Reasons for this include the high occurrence of false information during the 2016 US elections (Hughes & Waismel-Manor, 2021) and Brexit (Höller, 2021) as well as the COVID-19 infodemic (WHO, 2023). The transformation in the media and technology sector has created a range of new opportunities to manipulate people. The widespread adoption of social media applications has lowered the costs of information dissemination (Stieglitz & Dang-Xuan, 2013). Social media applications, platforms, and networks can be understood as sociotechnical systems that integrate technological with social components. Unlike traditional media, highly connected communication does not always provide journalists as gatekeepers of accurate information (Torre, 2022). The danger of unconscious influence by fake news increases with the development of artificial intelligence (Bastick, 2021). Recent progress in generative AI has become worrisome due to their proficiency in crafting human-like and persuasive content. Furthermore, their inherent propensity to create false information, hallucinated text, fake images and videos, exemplified by "deepfake" technology, is especially alarming. In discourse, attempts by organs of other states to manipulate citizens in internet media are even referred to as digital warfare (Zannettou et al., 2019). In combating all forms of FI, it is helpful to understand what makes people susceptible to it and what makes them believe misleading content (Pennycook & Rand, 2021). From a security perspective, belief in FI is critical when it functions as a human vulnerability that can be exploited. In sociotechnical systems, adversaries can leverage misinformation and disinformation to induce insecure actions (e.g., clicking malicious links, disclosing sensitive data), manipulate organizational decisions, or erode trust in security-relevant communications. Hence, belief in FI can be understood as a human-centered security risk that increases the attack surface of sociotechnical systems.

There are several definitions and terms in the field of FI. The term "misinformation" can be defined as unintended misleading representational content (Søe, 2018), while "disinformation" is described as misleading information that also has the function of misleading (Fallis, 2015). An often-used term is "fake news," which describes the intentional creation of pseudo-journalistic disinformation (Egelhofer & Lecheler, 2019). Since these terms are already associated with specific characteristics in the literature (e.g. intent to deceive and form of presentation), we refer to the general term "false information" (FI) from Walke & Nürnberger (2025): "untrue information that does not correspond to verifiable facts or reality". The term false information also refers to fake information (e.g. deepfakes), since the definition includes the non-existent reference to reality. In this paper, "security risk" refers to a socio-technical risk arising from human susceptibility that can be exploited to cause harm (e.g., decision manipulation, unauthorized disclosure, disruption of trusted communication). "Human-centered security" emphasizes that humans are integral components of security-critical systems and that cognitive and behavioral factors shape system security. A crucial factor regarding FI is the *belief* in the information truthfulness of the recipient. According to Hall (1973), messages are encoded by the producer and decoded by the recipient. Only the decoded meaning is effective and leads to cognitive, emotional, ideological, or behavioral consequences (Hall, 1973). We define belief in FI as the recipient's acceptance of untrue information as factual during the decoding process, where the perceived truthfulness of the decoded message—rather than its actual veracity—determines the cognitive, emotional, ideological, or behavioral consequences for the individual.

Previous reviews in the field of *FI belief* mainly considered a small number of studies and focused on fake news recognition, trust and engagement intentions (Bryanov & Vziatysheva, 2021). A comprehensive literature review and systematization of a larger amount of primary and high-quality studies regarding FI belief is still pending. This gap is addressed by the present paper with a comprehensive literature review, which is based on the framework from vom Brocke et al. (2009). This literature review deals with the following research question: "*Which factors influence the belief in false information?*" The objective of this review is to identify and categorize the most relevant factors that influence the belief in FI. Moreover, preventive strategies are explored to strengthen security in sociotechnical systems. In the next sections, we outline our method (2.), our findings (3.), and we discuss the findings, outline the limitations of this study and the implications for research and practice (4.).

## 2. Methodology

The literature review is based on the framework for literature reviewing from vom Brocke et al. (2009), which consists of five phases: definition of scope, conceptualization, literature search, analysis and synthesis, and research agenda. The scope of our literature review can be systematized referring to the taxonomy of Cooper (1988) as recommended by vom Brocke et al. (2009). Cooper (1988) subdivides literature reviews into six characteristics: focus, goal, perspective, coverage, organization, and audience of the paper which are further divided into categories (Table 1). The focus of this paper lies on research outcomes. Our primary goal is the integration of existing research results in the field of believing in false information. Our secondary goal is to show central issues regarding research in belief in false information. We choose a neutral perspective for the literature review, the presentation of the analysis and synthesis of the results.

**Table 1. Positioning of this paper within the taxonomy of Cooper (1988)**

|  | Categories | | | |
|---|---|---|---|---|
| Focus | **Research outcomes** | Research methods | Theories | Practices or applications |
| Goal | **Integration** | | Criticism | **Central issues** |
| Perspect. | **Neutral representation** | | Espousal of position | |
| Coverage | Exhaustive | Selective | **Representative** | Central or pivotal |
| Organizat. | Historical | | **Conceptual** | Methodological |
| Audience | **Specialized scholars** | **General scholars** | **Practitioners** | General public |

The coverage of the literature is representative and the organization of the papers is conceptual. Our target groups are specialized scholars, general scholars and practitioners (Table 1).

The conceptualization (vom Brocke et al., 2009) is based on previous research results in the field of FI and was primarily used to identify dimensions and relevant search terms to answer the research question. The literature search consisted of techniques of keyword search, backward and forward search in different sources (citation indexing services, publications and the database Business Source Ultimate, APA PsycInfo, Applied Science & Technology Source, Sociology Source Ultimate, ACM Digital Library). The search was filtered to literature that is peer-reviewed and published in academic journals. It was filtered for primary studies with a medium to high level of evidence, excluding studies such as single case observations and expert opinions. Search terms included "fake news", "misinformation", "disinformation", "rumor", "belief" and "believ*". The penultimate phase is the analysis and synthesis of the found literature, where a concept matrix approach was applied (Salipante et al., 1982; Webster & Watson, 2002).

The found influence factors were classified in four different categories (Table 2): *extensively examined*, *moderately examined*, *minimally examined*, and *contradictory results*. This classification was based on two principal criteria—the consistency of study results and the number of supporting publications. While effect sizes offer valuable quantitative insights, methodological differences may sometimes lead to notable variability. By considering both the consistency of findings and the extent of supporting evidence, our approach aims to capture patterns that are stable and generalizable. This dual-criterion method provides a measured and balanced synthesis of the evidence, even when direct comparisons of effect sizes are challenging.

**Table 2. Categories of influence factors**

| Extensively examined | ↑↑↑ | Consistent results in at least 3 and in more than 75% of the studies |
|---|---|---|
| Moderately examined | ↑↑ | Consistent results in at least 2 studies without contrary results |
| Minimally examined | ↑ | Studies that prove an impact predominate |
| Contradictory results | ↔ | Study results up to now show no clear direction of influence |

A factor was considered 'extensively examined' if at least 75% of the results across a minimum of three independent studies were consistent, indicating a strong empirical basis. If a factor showed 100% consistency, but was supported by only two studies, it was categorized as 'moderately examined'. While the findings in this category were fully aligned, the limited number of studies prevented a definitive classification as 'extensively examined', highlighting a possible need for further research.

'Minimally examined' was assigned to factors that demonstrated at least 66.6% consistency across a minimum of two studies, with a greater number of significant results compared to non-significant or contradictory ones. This category suggests an emerging but not yet fully established influence, requiring further validation. In contrast, factors with contradictory results showed no clear direction of influence, indicating inconsistencies in the existing research. Such discrepancies may arise due to methodological differences, sample variations, or external influencing variables.

This systematic approach ensures a structured evaluation of the existing literature by assessing both the reliability and the extent of empirical support for each factor. The classification facilitates a clear synthesis of findings and serves as a foundation for developing a research agenda. This research agenda, which is based on the framework from vom Brocke et al. (2009), was developed in the final phase through an analysis and synthesis of the literature results.

## 3. Findings

In the analysis and synthesis of this literature review, 24 influence factors were identified, which can be assigned to the following six main categories: demographic factors, personality traits and others, psychological factors, policy- and values-based factors, media consumption and preventive factors. A scientific consensus on their influence (highly relevant influence factor) was found for 16 factors concerning the belief in FI. We present the findings of this review successively based on the six main categories.

**Table 3. Demographic factors**

| Effect | Literature |
|---|---|
| **Age** | |
| Young (7) ↑ | Ahmed & Rasul, 2022; Gupta et al., 2023; Halpern et al., 2019, pp. 226–228; Pan et al., 2021; Pickles et al., 2021; Soetekouw & Angelopoulos, 2022; Wolverton & Stevens, 2020 |
| Old (1) | J. P. Baptista et al., 2021 |
| No effect (3) | Abraham & Mandalaparthy, 2021; Greenhill & Oppenheim, 2017; Koch et al., 2023 |
| **Gender ↔** | |
| Women (3) | Halpern et al., 2019, p. 228; Lai et al., 2020; Pan et al., 2021 |
| Men (2) | Filkuková et al., 2021; Pickles et al., 2021 |
| No effect (3) | Greenhill & Oppenheim, 2017; Gupta et al., 2023; Wolverton & Stevens, 2020 |
| **Education** | |
| Low education (11) ↑↑↑ | J. P. Baptista et al., 2021; A. Enders et al., 2023; Filkuková et al., 2021; Lai et al., 2020; Pan et al., 2021; Pickles et al., 2021; Qerimi & Gërguri, 2022; Rampersad & Althiyabi, 2020; Soetekouw & Angelopoulos, 2022; Wolverton & Stevens, 2020; Zrnec et al., 2022 |
| No effect (1) | Greenhill & Oppenheim, 2017 |
| **Economic situation** | |
| Less favored (2) ↑ | Faragó et al., 2020; Pan et al., 2021 |
| More favored (1) | Qerimi & Gërguri, 2022 |
| No effect (1) | Greenhill & Oppenheim, 2017 |

Among the demographic factors, only the influence of educational level studied is scientifically undisputed. Individuals with a low educational level are more likely to believe FI. Additionally, there is a tendency for younger people to believe FI more readily than older people. Table 3 provides an overview of all the demographic factors. It shows the number of analyzed results per factor studied, the symbols of the analysis results, and the extent of the positive influencing factors. The age factor appears to be a weak indicator, if at all. Firstly, the number of studies that find no significance is high. Secondly, Wolverton and Stevens (2020) describe the effect of age on the correct identification of fake news as minimal. For instance, Pan et al. (2021) report that the age of participants is negatively associated with the acceptance of misinformation (β = -0.05, p < 0.001). The results suggest that younger individuals are more likely to believe FI. Regarding gender, there seems to be no clear direction of influence. Only a few of the studied papers have explored the influence of economic situation on belief to FI. The four studies considered yield three different results, suggesting that the economic situation likely plays a minor or no role in the belief in FI. However, there is a tendency for economically disadvantaged individuals to be more willing to believe FI.

Table 4. Personality traits and other factors

| Effect | Literature |
| --- | --- |
| **Extraversion** | |
| **High extraversion (7)** ↑↑↑ | Ahmed & Rasul, 2022; Ahmed & Tan, 2022; Calvillo, Garcia, et al., 2021; Lai et al., 2020; Sindermann et al., 2021; Wolverton & Stevens, 2020; Zrnec et al., 2022 |
| Low extraversion (1) | Doughty et al., 2017 |
| **Openness** ↔ | |
| High openness (2) | Ahmed & Rasul, 2022; Wolverton & Stevens, 2020 |
| Low openness (2) | Calvillo, Garcia, et al., 2021; Doughty et al., 2017 |
| No effect (2) | Sindermann et al., 2021; Zrnec et al., 2022 |
| **Neuroticism** | |
| **High neuroticism (2)** ↑ | Ahmed & Rasul, 2022; Lai et al., 2020 |
| Low neuroticism (1) | Doughty et al., 2017 |
| **Agreeableness** | |
| **Low agreeableness (3)** ↑↑↑ | Ahmed & Tan, 2022; Calvillo, Garcia, et al., 2021; Doughty et al., 2017 |
| No effect | Zrnec et al., 2022 |
| **Conscientiousness** | |
| High conscient. (1) | Wolverton & Stevens, 2020 |
| **Low conscient. (4)** ↑↑↑ | Ahmed & Rasul, 2022; Calvillo, Garcia, et al., 2021; Doughty et al., 2017; Zrnec et al., 2022 |
| **Conspiracy mentality** | |
| **High conspiracy mentality (5)** ↑↑↑ | Anthony & Moulding, 2019; Calvillo, Rutchick, & Garcia, 2021; Frischlich et al., 2021; Halpern et al., 2019, p. 226; Szebeni et al., 2021 |
| **Analytical thinking** | |
| **Low cognitive reflection (13)** ↑↑↑ | Ahmed, 2021; Ahmed & Rasul, 2022; Ahmed & Tan, 2022; J. Baptista et al., 2021; Calvillo, Rutchick, & Garcia, 2021; Li et al., 2022; Nurse et al., 2022; Pennycook, McPhetres, et al., 2020; Pennycook & Rand, 2019; Rahmanian & Esfidani, 2022; Ross et al., 2021; Sindermann et al., 2021; Zrnec et al., 2022 |

Table 4 provides an aggregated summary of the results related to *personality traits*, while in addition to the big five personality traits, the analyzed literature frequently discussed *conspiracy mentality* and *analytical thinking*, which are described as 'others' in Table 4. *Extroverts* appear to be more likely to believe in FI than introverts. The personality trait of *openness* does not seem to have any influence. The results regarding the influence of *neuroticism* on FI belief are inconclusive. However, there is a tendency for individuals with high levels of neuroticism to believe in FI. There is a scientific consensus that individuals with low *agreeableness* are more likely to believe FI.

People with low *conscientiousness* are more likely to believe FI. All the analyzed studies conclude that a *conspiracy mentality* contributes to people more easily believing FI. The analysis included results on cognitive reflection in the category of analytical thinking, typically measured using a cognitive reflection test (CRT), as well as results on intelligence and analytical thinking. Low cognitive reflection is confirmed as an influencing factor for the belief in FI.

Table 5 deals with factors from the field of *psychology*. It examines the influence of emotion, fear, exposure and repetition, and the confirmation bias on FI belief. Table 5 shows that all considered factors are positively correlated with belief in FI. When people rely on *emotions* to help distinguish between true and false information, it increases the likelihood that they will believe fake news (Martel et al., 2020). Negative emotions, in particular, appear to be positively associated with the belief in FI (Li et al., 2022; Porter et al., 2003). *Fear* positively influences the belief in FI. The analyzed literature examined fear of missing out (FoMO), fears of anti-vaxxers, and general health-related fears.

Table 5. Psychological factors

| Effect | Literature |
| --- | --- |
| **Emotion** | |
| **Positive correl. (6)** ↑↑↑ | Bago et al., 2022; Bonnin & Sinno, 2022; Li et al., 2022; Martel et al., 2020; Nahleen et al., 2021; Porter et al., 2003 |
| **Fear** | |
| **Positive correl. (4)** ↑↑↑ | Ali et al., 2022; Freiling et al., 2023; Pan et al., 2021; Pundir et al., 2021 |
| No effect (1) | Talwar et al., 2019 |
| **Exposure and repetition** | |
| **Positive correl. (5)** ↑↑↑ | Ahmed, 2021; Greenhill & Oppenheim, 2017; Pan et al., 2021; Pennycook et al., 2018; Polage, 2012 |
| No effect (1) | Nahleen et al., 2021 |
| **Confirmation bias** | |
| **Positive correl. (6)** ↑↑↑ | Horner et al., 2021; Kim et al., 2019; Koch et al., 2023; Moravec et al., 2019; Pan et al., 2021; Tsang, 2021 |

The results of the reviewed literature regarding the influence of *exposure and repetition* on the belief in FI are consistent. Exposure and repetition reinforce the belief in FI. Exposure to deepfakes leads to a perceived accuracy increase of β = 0.185 at p < 0.05 (Ahmed, 2021). Another psychological effect discussed and investigated in the relevant research is confirmation bias. *Confirmation bias* suggests that people are more likely to believe information that aligns with their pre-existing opinions (Housholder & LaMarre, 2014). All analyzed studies confirm that confirmation bias is a factor influencing belief in FI.

**Table 6. Policy- and value-based factors**

| Effect | Literature |
|---|---|
| **Alignment with political orientation** | |
| Positive correlation (8) ↑↑↑ | Anthony & Moulding, 2019; Faragó et al., 2020; Michael & Breaux, 2021; Moravec et al., 2019; Pereira et al., 2023; Reedy et al., 2014; Szebeni et al., 2021; Turel & Osatuyi, 2021 |
| **Value-based** | |
| Conservative (5) ↑↑↑ | Calvillo, Rutchick, & Garcia, 2021; Gupta et al., 2023; Horner et al., 2021; Ognyanova et al., 2020; Samore et al., 2018 |
| **Political orientation** | |
| Right-wing (4) ↑↑↑ | Arendt et al., 2019; J. P. Baptista et al., 2021; Halpern et al., 2019, p. 226 |
| No effect (1) | Frischlich et al., 2021 |

Table 6 shows *policy- and value-based influence factors*. *Alignment with political orientation* has a positive effect on belief in FI in all studied cases. Eight studies confirm the positive effect on the belief in FI. All analyzed studies consider people with conservative values to be more susceptible to FI belief. The analyzed literature concludes that right-leaning individuals are generally more susceptible to FI.

**Table 7. Media consumption**

| Effect | Literature |
|---|---|
| **Usage of social media** | |
| Positive correlation (2) ↑ | Jamieson & Albarracín, 2020; Su, 2021 |
| Low usage of social media (1) | Halpern et al., 2019 |
| No effect (1) | A. Enders et al., 2023 |
| **News consumption in social media** | |
| Positive correlation (4) ↑↑↑ | Ahmed & Rasul, 2022; A. M. Enders et al., 2021; Jamieson & Albarracín, 2020; Stecula & Pickup, 2021 |
| **News consumption ↔** | |
| High news consumption (1) | Calvillo, Garcia, et al., 2021 |
| Less news consumption (1) | Filkuková et al., 2021 |
| News consumption as resilience factor (1) | Altay et al., 2023 |

In Table 7 factors regarding *media consumption* are presented. The analyzed literature frequently includes results on general news consumption, news consumption on social media, and social media usage. There are no definitive results indicating that the *use of social media* makes individuals particularly believe in FI. There are only tendencies suggesting this might be the case. *News consumption on social media* makes people believe in FI. The influence on the belief in false information is confirmed by four studies. Regarding *general news consumption,* three analyzed studies yield contradictory results.

**Table 8. Preventive factors**

| Effect | Literature |
|---|---|
| **Labeling of false information** | |
| Lack of labeling (13) ↑↑↑ | Arendt et al., 2019; Clayton et al., 2020; Ecker et al., 2020; Garrett & Poulsen, 2019; Grady et al., 2021; Hernandez-Mendoza et al., 2022; Kirchner & Reuter, 2020; Koch et al., 2023; Lee, 2022; Mena, 2020; Moravec et al., 2020; Pennycook, Bear, et al., 2020; Pennycook et al., 2018 |
| No effect (1) | Moravec et al., 2019 |
| **Thinking about correctness** | |
| Lack of reflection (2) ↑↑ | Lewandowsky & van der Linden, 2021; Pennycook, McPhetres, et al., 2020 |
| **Knowledge and training** | |
| Lack of knowledge in FI recognition (5) ↑↑↑ | Apuke et al., 2023; Green et al., 2022; Guess et al., 2020; Lutzke et al., 2019; Soetekouw & Angelopoulos, 2022 |
| No effect (1) | Badrinathan, 2021 |

In Table 8 *preventive factors* are examined. Commonly studied preventive factors include labeling FI, prompting individuals to think about the accuracy of the information, and knowledge or training. When a factor proved effective, the absence of this measure was considered a factor influencing belief in FI. All preventive factors essentially showed an effect. The consolidated results are presented in Table 8.

The effectiveness of *labeling FI* has been confirmed in the studied research. The belief in FI is reduced by a warning label on the information. Individuals who *reflect on the correctness* of information (e.g., after a prior general warning about false information) are less likely to believe FI. Individuals who have *knowledge* of how to recognize FI are less susceptible to it. Therefore, they believe FI less than those without this knowledge.

## 4. Discussion and Conclusion

This comprehensive literature review provides an in-depth analysis of the factors influencing belief in FI. The findings reveal several critical insights into why individuals believe FI. This review identified 24 factors in total and 16 highly relevant influence factors contributing to the belief in FI. The results of this review show that the majority of factors arise from social sciences and psychology.

The review indicates that individuals with lower education levels are more prone to believing FI. This highlights a significant gap in media literacy and critical thinking skills among certain educational demographics. Addressing this gap through educational interventions, such as incorporating media literacy courses in school curriculums and adult education programs, could empower individuals to critically evaluate information. While age and gender showed less consistent effects, the tendency for younger individuals to be more susceptible to FI suggests a need for targeted interventions for this demographic, possibly through platforms and channels they frequent.

Personality traits significantly impact the belief in FI. High extraversion, low agreeableness, and high neuroticism are consistently linked to a greater tendency to believe FI. Extraverted individuals may be more exposed to varied information sources due to their social nature, increasing their chances of encountering FI. Low agreeableness might correlate with a higher likelihood of questioning mainstream information but also being susceptible to alternative narratives. High neuroticism could make individuals more vulnerable to FI due to heightened anxiety and emotional instability. Promoting emotional resilience and critical reflection could mitigate these effects.

Emotions, fear, exposition and repetition, and confirmation bias are potent psychological drivers of FI belief. Emotional reasoning can cloud judgment, leading to susceptibility to FI. Negative emotions, particularly fear, can increase the likelihood of accepting FI, as fear-based narratives are compelling and memorable. Repeated exposure to the same misinformation can reinforce belief through the mere-exposure effect. Confirmation bias, where individuals favor information aligning with their pre-existing beliefs, further entrenches false beliefs. Interventions must focus on these psychological aspects, perhaps through public awareness campaigns that highlight these biases and promote critical thinking.

Alignment with political orientation and conservative values emerged as significant predictors of FI belief. Ideological biases play a crucial role in how individuals evaluate information. This suggests that combating FI is not just about correcting facts but also addressing the underlying ideological contexts. Tailored communication strategies that consider these ideological dimensions could be more effective. For instance, engaging trusted community leaders or influencers within specific ideological groups can help disseminate accurate information more credibly.

Social media usage is linked to increased belief in FI, though the results are not entirely conclusive. The format and algorithms of social media platforms, which might prioritize engagement over accuracy, could amplify FI. Personalized filter algorithms of social media allow users to find confirmation of their personal beliefs and opinions in echo chambers (Wardle & Derakhshan, 2017). Moreover, the journalistic "gatekeeper" role regarding the accuracy of information is of secondary importance on social media (Torre, 2022). News consumption on social media platforms correlates with higher FI belief, indicating that the source and context of news consumption are critical. Traditional media, while not immune to spreading FI, often has more checks and balances compared to social media. Encouraging critical media consumption habits, such as verifying sources and cross-referencing information, could help mitigate the impact of FI.

Labeling FI, promoting reflection about correctness, and improving knowledge and training can be effective strategies in reducing FI belief. Labels warning about the potential falsity of information can create a cognitive pause, prompting individuals to reconsider their initial reactions. However, such labels also have undesirable side effects. For instance, there is a risk that information not labeled as false is more likely to be considered true (Pennycook, Bear, et al., 2020). Labels may also lead users to trust flags blindly, even if the assessment is inaccurate (Gaozhao, 2021; Lu et al., 2022). Encouraging correctness reflection through prompts and educational tools can help individuals develop a habit of critically assessing information. Knowledge and training programs that enhance individuals' ability to recognize FI are crucial. These could include workshops, online courses, and interactive media that teach critical evaluation skills.

The preventive factors identified in this review can be interpreted as human-centered security controls. Warning labels and accuracy prompts introduce friction that disrupts automatic acceptance, while training increases the capability to detect deception. However, these controls may introduce security-relevant side effects (e.g., overreliance on flags or implied-truth effects), which must be considered in platform and organizational security design.

This review has limitations that need to be taken into account. This review does not analyze technical attack vectors or system compromise directly; instead, it focuses on belief formation as a human vulnerability relevant to security in sociotechnical systems. Additionally, the fact that part of the analyzed literature was found through snowball sampling might have led to an imbalance in the research areas studied. The studies tend to highlight significant core findings. Factors that do not show significance are sometimes not discussed or only mentioned in passing, making them easy to overlook in the literature analysis and thus not included in the analyses. This could lead to publication bias, making some factors appear more significant than they actually are. Although the total number of analyzed results is high, the results per examined factor are sometimes low due to the abundance of factors and the availability of studies on sub-areas. In future research, more meaningful results could be achieved by examining individual factors in a targeted manner with a higher hit rate. Future research should continue to explore these factors in greater detail, considering the evolving nature of technology and media landscapes. Longitudinal studies could provide insights into how belief in FI changes over time and in response to different interventions. Additionally, cross-cultural studies could highlight how different societal contexts influence susceptibility to FI, helping to tailor interventions to specific cultural and societal needs.

This literature review underscores the multifaceted nature of belief in FI, influenced by a combination of demographic, personality, psychological, ideological, and media consumption factors. The insights gained from this review highlight the need for a holistic approach to combat fake news, misinformation, disinformation or fake information, for example, deepfakes generated by artificial intelligence. Educational programs that foster critical thinking and media literacy, coupled with targeted interventions that address emotional and psychological biases, are essential. Policymakers and educators should focus on creating environments that promote analytical thinking and skepticism towards unverified information. Furthermore, social media platforms must implement robust measures to label and mitigate the spread of FI, considering the significant role of repeated exposure and confirmation bias. The review identifies key preventive factors such as labeling false information, encouraging reflective thinking, and enhancing knowledge and training to reduce FI belief. These strategies can be integrated into broader educational and policy frameworks to build societal resilience against FI.

By understanding the diverse factors that contribute to FI belief, stakeholders can develop more effective strategies to protect individuals and society from the detrimental effects of false information. This holistic understanding is crucial in an era where the rapid dissemination of information, facilitated by technological advancements, poses significant challenges to discerning truth from falsehood.